\documentclass[apj]{emulateapj}

\usepackage{graphicx}
\usepackage{bm}
\usepackage[dvips]{color}

\begin{document}

\title{Comparison of statistical treatments for
the equation of state for\\ core-collapse supernovae}

\author{S.~R. Souza\altaffilmark{1},
A.~W. Steiner\altaffilmark{2},
W.~G. Lynch\altaffilmark{2},
R. Donangelo\altaffilmark{1},
M.~A. Famiano\altaffilmark{3}}

\altaffiltext{1}{Instituto de F\'\i sica, Universidade Federal do
Rio de Janeiro Cidade Universit\'aria, CP 68528, 21941-972, Rio de
Janeiro, Brazil and
Instituto de F\'\i sica, Universidade Federal do Rio Grande do Sul
Av. Bento Gon\c calves 9500, CP 15051, 91501-970, Porto Alegre,
Brazil}

\altaffiltext{2}{Joint Institute for Nuclear Astrophysics,
National Superconducting Cyclotron Laboratory, and
the Department of Physics and Astronomy, Michigan State University,
East Lansing, MI 48824, USA}

\altaffiltext{3}{Physics Department, Western Michigan
University, Kalamazoo, MI 49008, USA}

\begin{abstract}
 Neutrinos emitted during the collapse, bounce and subsequent
 explosion provide information about supernova dynamics. The neutrino
 spectra are determined by weak interactions with nuclei and nucleons
 in the inner regions of the star, and thus the neutrino spectra are
 determined by the composition of matter. The composition of stellar
 matter at temperature ranging from $T=1-3$ MeV and densities ranging
 from $10^{-5}$ to $0.1$ times the saturation density is
 explored. We examine the single-nucleus approximation
 commonly used in describing dense matter in supernova simulations and
 show that, while the approximation is accurate for predicting the
 energy and pressure at most densities, it fails to predict the
 composition accurately. We find that as the temperature and density
 increase, the single nucleus approximation systematically
 overpredicts the mass number of nuclei that are
 actually present and underestimates the contribution from lighter
 nuclei which are present in significant amounts.
\end{abstract}


\section{\label{sec:introuduction}Introduction\protect}

Stars with masses larger than about 10 M$_{\odot}$ end their lives in
core-collapse supernovae. The initial collapse is initiated by the
disappearance of the pressure support driven by (i) dissociation of
nuclei and (ii) electron capture on nuclei which make matter
increasingly neutron rich. The electron captures emit neutrinos, which
initially escape, but are later trapped by the increasing density in
the core. The collapse is halted by the strong repulsion from high
density matter and the core ``bounces,'' driving an outward shock
wave. The standard view is that this shock loses energy through
continuing dissociation of nuclei and neutrinos from the core finally
restore energy to the shock driving an explosion.

The evolution of the supernova is determined by the equation of state
(EOS) and composition of matter at densities up to the nuclear
saturation density, $\rho_0 \sim 2.7 \times 10^{14}~\mathrm{g/cm}^3$
and at temperatures between 1 and 3 MeV. The collapse is driven by the
tendency of matter to have a low entropy per baryon~\citep{Bethe79},
the extent of electron captures during collapse is determined by the
composition~\citep{Hix03}, and the neutrino
spectrum~\citep{Sumiyoshi05} is determined by the nature of matter at
the ``neutrinosphere'', the surface of the last neutrino
  scattering, which is typically at $10^{11}$ g/cm$^3$.

At densities below $\rho \sim 8 \times 10^{13}~\mathrm{g/cm}^3$, most
matter resides in nuclei with $A\geq10$ arranged in a Coulomb
lattice, surrounded by a nucleon gas and embedded in a degenerate
electron gas. Masses of isolated nuclei in coexistence with the gas
may exceed those of stable nuclei but are limited to $A<1000$ by the
interplay of the Coulomb and symmetry contributions to their binding
energies. Both nucleons and nuclei together compose a nuclear (Fermi)
liquid-gas phase mixture, whose Equation of State (EOS) contributes
significantly to the EOS for supernova simulations.

A principal approach to computing the EOS for supernovae used in the
past three decades has been the ``Single-Nucleus Approximation'' (SNA)
where low-temperature matter is assumed to be composed of neutrons,
protons, alpha particles, and a single, representative, heavy
nucleus \citep{Lattimer85,LattimerSwesty,Shen98}.
The nuclei vaporize at temperatures that depend
sensitively on density. \citet{BurrowsLattimer} have found that the
SNA can accurately predict thermodynamic functions for the matter, and
microscopic calculations for the nuclear matter
EOS have been performed in this approximation for densities ranging
from 0 to saturation density \citep{Lattimer85,LattimerSwesty,Shen98}.
However, recent studies suggest that the SNA may be insufficient to
completely describe the dynamics of core-collapse supernovae. In
particular, \citet{Hix03} find that calculating electron captures on a
distribution of nuclei in nuclear statistical equilibrium, in place of
a single representative nucleus, increased the potential for an
explosion. Also, \citet{Arcones08} find that the presence
of light nuclei is important for describing the neutrino
spectra in the early post-explosion phase.

A more realistic description of supernovae EOS would model the
composition of matter by an ensemble of nuclei in near nuclear
statistical equilibrium. In this work, we refer to this approach as
the ``Grand Canonical Approximation'' (GCA). The GCA permits
investigation of the shapes and widths of the nuclear mass and charge
distributions and the population of nuclear excited states. In
situations where the Wigner-Seitz approximation for a unit cell is
valid and the widths of the heavy nuclear distributions can be
neglected, GCA and SNA models should make similar predictions. In a
recent paper \citet{BotvinaSNova2005} report on GCA calculations for
supernovae and suggest that differences between their GCA and previous
SNA calculations should be important. However, they did not directly
compare SNA and GCA approximations to quantify the differences.

To address this question, we adopt a GCA model that is formally
equivalent to that of \citet{BotvinaSNova2005}, and construct an
analagous SNA model for comparisons to the GCA model. This
corresponding SNA model employs the same mass formula and level
densities that are used in a corresponding GCA model. We compare
predictions of the GCA model and its companion SNA model for two
different mass formulae, which differ by their treatment of the
nuclear surface symmetry energy. Then, we construct another SNA model
(ISNA) which is more similar to the modern tabulated EOS
constructed in \citet{LattimerSwesty} and is frequently employed in
supernovae simulations. The ISNA model includes interactions and the
quantum statistics of the nucleons in the gas, a more complete
description of the Coulomb energy, surface energy term which
vanishes in the small proton number limit, a critical temperature
which depends on the electron fraction, and effects associated with
the presence of a neutron skin for nuclei with $N>Z$.

From comparisons between the GCA and SNA models, we find that the
traditional SNA approach succeeds at most densities in describing the
basic properties of the equation of state, but that it fails in
describing the composition. We find that the differences between the
predictions of GCA models and SNA models exceeds the differences
between the predictions of the various SNA models. More specifically,
we find that average quantities, such as the mass fractions for
neutrons, $\alpha$ particles and heavy nuclei, energy/baryon, and
entropy, are similar at $\rho\leq0.01$ and $Y_e=0.4$ in the GCA and
SNA approaches. The agreement is slightly worse for $Y_e=0.2$, and the
agreement between GCA and SNA deteriorates even more
for calculations that neglect the surface symmetry energy.

We find generally that the SNA and ISNA models agree quite well
over most of the densities and temperatures considered; the largest
differences are a few tenths of an MeV in the energy per baryon at
higher densities and comparable differences in the entropy. The SNA
and ISNA calculations, however, neglect completely the
considerable widths of the mass and charge distributions predicted by
the GCA calculations. At higher temperatures and densities, the SNA
calculations systematically over-predict the size of the
representative nucleus, and fail to treat the light nuclei that
become more abundant in some regimes of density and temperature. These
differences can have an impact on the electron capture and weak
interaction rates that prevail in a supernova.

In the following, we begin our comparisons by adopting identical
formula for the masses, level densities, and electron screening
approximations for pairs of GCA and SNA calculations. We perform this
comparison for two liquid drop models, LDM1, which has a
surface symmetry energy term, and LDM2, which does not. Section 2
describes these model assumptions. As these two calculations neglect
the interactions and quantum statistics of particles in the gas, the
ISNA model that contains these effects is described in Section 2 as
well. Section 3 describes and compares results obtained for the
various GCA and SNA models. Section 4 summarizes this work and
provides an outlook for future studies.

\section{\label{sec:models}The models\protect}

Sections 2.1-2.3 describe the SNA and GCA models without the inclusion
of the interactions and quantum statistics of particles in the gas.
These two calculations can be compared to isolate the differences
between the SNA and CGA models. Section 2.4 describes an improved
SNA model (the ISNA model) that includes the interactions and quantum
statistics of particles in the gas. This model allows tests of the
relative importance of some simplifying assumptions used in the models
described in Sections 2.1-2.3.

\subsection{\label{sec:groundState} Ground State Properties of Nuclei}

The binding energies of nuclei strongly influence the charge and mass
distributions of nuclei within hot extended nuclear systems
\citep{ISMMmass,ISMMlong}. In this work we use a form for the liquid
drop mass formula from \citep{massFormula}:

\begin{equation}
B_{A,Z}=C_v A - C_s A^{2/3} - C_c\frac{Z^2}{A^{1/3}}+\delta_{A,Z}A^{-1/2}
+C_d\frac{Z^2}{A}\;,
\label{eq:ldm}
\end{equation}

\noindent
where $A$ and $Z$ denote, respectively, the mass and atomic numbers,

\begin{equation}
C_i=a_i\left[1-k_i\left(\frac{A-2Z}{A}\right)^2\right]\;,
\label{eq:cldm}
\end{equation}

\noindent
and $i=v,s$ corresponds to volume and surface, respectively. In
\citet{ISMMmass}, its parameters have been fitted to the available
experimental data \citep{AudiWapstra}. The corresponding values are
listed in Table~\ref{tab:ldmpars}, and are labeled LDM1 (Liquid Drop
Model). For completeness, the pairing term is also included in the
expression above, although it is neglected in the calculations
presented in this paper. It reads:

\begin{equation}
\delta_{A,Z}=\left\{
   \begin{array}{rl}
      +C_p, & N\, {\rm and}\, Z\, {\rm even}\\
      0, & A\,\, {\rm odd}\\
      -C_p, & N\, {\rm and}\, Z\, {\rm odd}\\
   \end{array}\right.\;,
\label{eq:pairing}
\end{equation}

\noindent
where $N=A-Z$ stands for the number of neutrons. Besides the standard
terms, this parametrization includes a correction to the Coulomb
energy due to the diffuseness of the surface, $C_dZ^2/A$, as well as a
surface contribution to the symmetry energy. Both corrections are
usually neglected in simple parametrizations. While the latter
contribution is most important for light nuclei, it also influences
the masses of very heavy nuclei, as discussed below.

\begin{deluxetable*}{cccccccc}
\tablenum{1}
\tablecolumns{8}
\tablecaption{Parameters of the liquid drop mass formulas used in
this work. All the values are given in MeV.
\label{tab:ldmpars}}
\tablehead{
\colhead{Label} & \colhead{$a_v$} &
\colhead{$a_s$} & \colhead{$C_c$} &
\colhead{$C_d$} & \colhead{$C_p$} &
\colhead{$a_v k_v$} & \colhead{$a_s k_s$}}
\startdata
LDM1 & 15.6658 & 18.9952 & 0.72053 & 1.74859 & 10.857 & 27.7976 & 33.7053\\
LDM2 & 15.2692 & 16.038 & 0.68698 & 0.0 & 11.277 & 22.3918 & 0.0\\
\hline
 & $n_s$ & $n_d$ & ${\cal C}$ & $\zeta$ & $\sigma$ & $b$ & \\
\hline
LDM3 & 0.1764 & -0.2832 & 0.8990 & 0.9467 & 1.179 & 9.130 & \\
\enddata
\end{deluxetable*}

\begin{figure}
\vspace*{1.0cm}
\includegraphics[height=8.5cm,angle=-90]{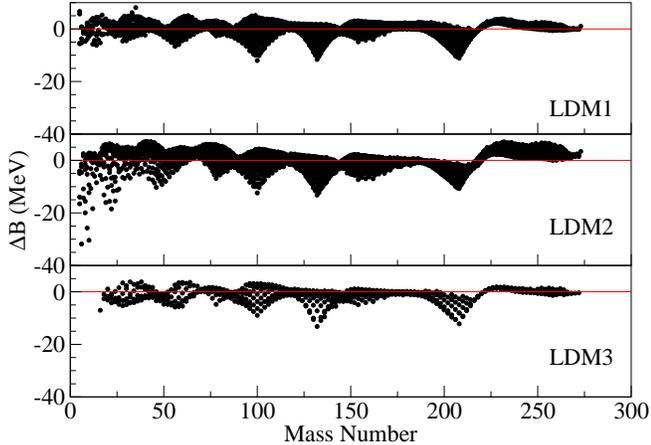}\break\break
\caption{\label{fig:diffBE}Difference between the total binding
energy predicted by the mass formula and the empirical values.
For details, see the text.}
\end{figure}

The accuracy of this formula can be inferred from
Figure~\ref{fig:diffBE}, which shows the difference $\Delta B$ between
the predictions of equation (\ref{eq:ldm}) and the empirical (measured)
values.  It is important to stress that $\Delta B$ corresponds to the
difference between the total binding energies, {\it i.e.}, it is not
divided by the mass number. Since the above formula has no corrections
associated with shell effects, pronounced discrepancies are observed
near closed shells. Shell effects for nuclei with $A>5$ are neglected
since they do not strongly modify the qualitative shape of the nuclear
distributions which we calculate.

In order to investigate the influence of the Coulomb diffuseness
correction and the surface symmetry energy, we have refitted the data
while keeping the parameters $C_d$ and $k_S$ equal to zero. In this
limit, this mass formula has no surface symmetry energy or diffuseness
correction to the Coulomb interaction, which makes it have the same
form that is used in many statistical calculations
\citep{BotvinaSNova2005,BotvinaSNova2004,smmIsobaric,Subal1999,Bondorf1995}.
The best fit parameters are listed in Table~\ref{tab:ldmpars} and
labeled LDM2. The corresponding deviations $\Delta B$ from the
empirical values are shown in the middle panel of
Figure~\ref{fig:diffBE}. One can clearly see tendencies for LDM2 to over
predict the empirical values for heavy masses and even stronger
tendencies to under predict the empirical values for light masses.

To aid in the discussion we define the nuclear symmetry energy,
$B_{\rm sym}$:
\begin{eqnarray}
B_{\rm sym} &=&
-\left(a_v k_v - \frac{a_s k_s}{A^{1/3}}\right)\frac{(A-2Z)^2}{A}\nonumber\\
& \equiv& -a_{\rm sym}\frac{(A-2Z)^2}{A}\; (\mathrm{for}~\mathrm{LDM2}).
\label{eq:symEnergy}
\end{eqnarray}
\noindent The second line shows the simplification that occurs by
setting $k_s=0$ in the LMD2 parametrization.

The difference between the LDM1 and LDM2 reflects the reduction of the
symmetry energy in the nuclear surface and in the effective symmetry
energy coefficient $a_{\rm sym}$ in the LMD1 parametrization, which
causes its masses to increase as a function of $A$, better following
the trends of the measured masses. In the simpler LDM2
parametrization, this coefficient is constant, which does not
accurately follow the experimental trends and leads to the systematic
deviations in both the light and heavy mass regions displayed in the
figure. We have checked that the influence of the $C_d$ coefficient
does not account for the increase in $\Delta B$ observed in the heavy
mass region.

Experimental values are used for the very light nuclei ($A < 5$),
whenever known, since both parameter sets give a very poor description
of the binding energies of light nuclei. This procedure is not adopted
for heavier nuclei as one has to consider nuclei far from the known
mass region. In this case, a careful extrapolation scheme should be
devised so as not to introduce spurious effects in the calculated
yields for mass regions in which no experimental information is
available \citep{ISMMmass,ISMMlong}. Since the major thrust of this
paper concerns the comparison of two approximations for calculating
equilibrium distributions of nuclei, the use of a mass formula for all
nuclei with $A > 4$ is sufficient and will ensure a smooth behavior
for the predicted yields throughout the mass range considered below.

A similar procedure is also adopted for the spin degeneracy factors of
nuclei. Empirical values are used only for $A<5$. For heavier nuclei,
we set the spin degeneracy factors to unity (vanishing nuclear
spin). This approximation is not that important because
errors introduced by neglecting ground state spin degeneracies are
much smaller than the uncertainties in the nuclear level densities at
finite temperatures. These level densities are described below.

\subsection{\label{sec:gc} The Grand Canonical Approximation (GCA)}

For a system composed of $\{k_1,k_2,\cdots,k_M\}$ different species
in thermal equilibrium at temperature $T$, the grand-partition
function is given by

\begin{eqnarray}
Z_G=& &\sum_{k_1=0}^\infty\cdots\sum_{k_M=0}^\infty Z_C(T,V,k_1,k_2,\cdots,k_M)
\nonumber\\
& &e^{\beta(\mu_1 k_1+\mu_2 k_2+\cdots+\mu_M k_M)}\;,
\label{eq:zg}
\end{eqnarray}

\noindent
where $\beta\equiv 1/T$, and the canonical partition function $Z_C$
is given by

\begin{eqnarray}
Z_C=& &\frac{\left[g_1 V_f A_1^{3/2}/\lambda_T^3\right]^{k_1}}{k_1!}
\cdots
\frac{\left[g_M V_f A_M^{3/2}/\lambda_T^3\right]^{k_M}}{k_M!}\nonumber\\
& & \times e^{-\beta F(T,V,k_1,\cdots,k_M)}\;.
\label{eq:zc}
\end{eqnarray}

\noindent
In the above expression, $\lambda_T=\sqrt{2\pi\hbar^2/mT}$, we
approximate the nucleus translational effective mass by
$m_{A,Z}=Am$ following \cite{ISMMlong}. We further approximate $m$ by
the free nucleon mass $m=939$~MeV/$c^2$. The function
$F(T,V,k_1,\cdots,k_M)$ denotes the Helmholtz free energy of the
system, excluding contributions associated with the translational
motion, and $g_i$ denotes the spin degeneracy factor. The free volume
$V_f$ takes finite size effects into account and is calculated as in
\citet{BotvinaSNova2004}, {\it i.e.}, $\Omega\equiv V_f/V =
1-\rho_B/\rho_0$, where $V$ is the total volume, $\rho_B$ denotes the
baryon density, and $\rho_0$ represents normal (saturation) nuclear
density. If $F(T,V,k_1,\cdots,k_M)$ can be written as

\begin{equation}
F=k_1f_1(T,V)+k_2f_2(T,V)+\cdots+k_Mf_M(T,V)\;,
\label{eq:fec}
\end{equation}

\noindent
the canonical partition function becomes

\begin{equation}
Z_C(T,V,k_1,\cdots,k_M)=\prod_{i=1}^M
\frac{\left[g_i V_f A_i^{3/2}/\lambda_T^3\right]^{k_i}}{k_i!}
e^{-\beta f_i k_i}\;,
\label{eq:zcf}
\end{equation}

\noindent
which leads to

\begin{equation}
Z_G(T,V,\{\mu_i\})=\prod_{i=1}^M e^{q_i}\;,
\label{eq:zgcf}
\end{equation}

\noindent
where

\begin{equation}
q_i=\frac{g_i V_f A_i^{3/2}}{\lambda_T^3}
e^{\beta (\mu_i-f_i)}\;.
\label{eq:qi}
\end{equation}

\noindent
In this case, the average number of nuclei of species $i$ in the
volume $V$, $Y_i$, can be easily evaluated

\begin{equation}
Y_i=\frac{1}{Z_G}
\sum_{k_1=0}^\infty\frac{q_i^{k_1}}{k_1!}\cdots
\sum_{k_i=0}^\infty k_i\frac{q_i^{k_i}}{k_i!}\cdots
\sum_{k_M=0}^\infty\frac{q_M^{k_M}}{k_M!}=q_i\;.
\label{eq:yi}
\end{equation}

By assuming chemical equilibrium, the chemical potentials may be written
as

\begin{equation}
\mu_{A,Z}=\mu_B A + \mu_Q Z,
\label{eq:chempot}
\end{equation}

\noindent so that the number density of a given species $(A,Z)$ becomes

\begin{equation}
n_{A,Z}=\frac{q_{A,Z}}{V}=\frac{g_i\Omega A^{3/2}}{\lambda_T^3}
e^{(\mu_B A + \mu_Q Z-f_{A,Z})/T}\;.
\label{eq:rhoaz}
\end{equation}

\noindent
Thus, $\mu_B$ and $\mu_Q$ are determined upon fixing the average number of
baryons and by imposing charge neutrality, respectively:

\begin{eqnarray}
n_B=\sum_{A,Z}A\,n_{A,Z}
\label{eq:rhob}\\
n_e=\sum_{A,Z}Z\,n_{A,Z}\;,
\label{eq:rhoc}
\end{eqnarray}

\noindent
where $n_e$ denotes the electron density. As in
\citet{BotvinaSNova2004}, the sums include all nuclei with
$Z\le A$. We have checked that, for the temperatures and densities
considered below, $A= 1000$ is a safe upper cut in the sums above. We
impose that the relative errors in equations~(\ref{eq:rhob}-\ref{eq:rhoc})
are smaller than $10^{-7}$.

In order to determine the chemical potentials, and subsequently, other
relevant thermodynamical quantities, the Helmholtz free energy
$f_{A,Z}$ must be specified. We include contributions to the internal
excitation of nuclei associated with surface and bulk, besides the
binding energy of the nuclei and the Coulomb interaction among the
particles:

\begin{eqnarray}
f_{A,Z} &=&\beta(T)A^{2/3}-\frac{T^2}{\epsilon_0}A-B_{A,Z}
\nonumber\\
& + & C_C\frac{Z^2}{A^{1/3}}\left[-\frac{3}{2}
\left(\frac{n_B}{n_0}\right)^{1/3}
+\frac{1}{2}\left(\frac{n_B}{n_0}\right)\right]\;,
\label{eq:fazgc}
\end{eqnarray}

\noindent
where

\begin{equation}
\beta(T)=\beta_0\left[\left(\frac{T_c^2-T^2}{T_c^2+T^2}\right)^{5/4}-1\right]
\;.
\label{eq:beta}
\end{equation}

\noindent The values of the parameters $\beta_0=18.0$~MeV,
$T_c=18.0$~MeV, and $\epsilon_0=16.0$~MeV correspond to those usually
adopted in statistical calculations
\citep{smmIsobaric,BotvinaSNova2004,ISMMmass,ISMMlong,Bondorf1995}.
The Coulomb energy, in the last term in equation~(\ref{eq:fazgc}), is
calculated through the Wigner-Seitz approximation \citep{WignerSeitz}.
It should be noted that equation~(\ref{eq:fazgc}) differs from the
usual expressions employed in statistical models because it takes into
account the presence of the electron gas surrounding the nuclei
\citep{BBPsupernova}. In this respect, an important comment is in
order. In principle, the baryon density in equation~(\ref{eq:fazgc})
should be the average particle density in the Wigner-Seitz cell
\citep{BBPsupernova}. However, if this is done, the Helmholtz free
energy cannot be cast in the form of equation~(\ref{eq:fec}) and,
then, the simple formulae above would no longer be valid. Thus, as in
\citet{BotvinaSNova2004}, we approximate the average density by $n_B$
in order to keep the simple relations above. Since we want to maintain
the GCA and the SNA close as possible, we adopt this approximation in
SNA model as well.

Finally, nuclei with $A<4$ are treated as point particles, with no
internal degrees of freedom. Therefore, the temperature dependent
terms in equation~(\ref{eq:fazgc}) are set to zero for those nuclei.

\subsection{\label{sec:singNuc} The Single Nucleus Approximation (SNA)}

Since the determination of the chemical potentials above and the
subsequent computation of the relevant thermodynamical quantities is
too time consuming to be used in many practical astrophysical
calculations, the GCA has been simplified
\citep{SingleNucleus1975,BurrowsLattimer,Lattimer85,LattimerSwesty}.
More specifically, instead of considering all possible species, only a
few types are allowed: neutrons, protons, alpha particles and a single
species of heavy nucleus. The latter should, to some extent, represent
the nucleus distribution at the heavy mass region. The other three
nuclei are intended to consider the main contribution to the nucleus
distribution in the light mass region.

The constraints associated with baryon number and charge neutrality
simplify to

\begin{equation}
n_B=n_n+n_p+4n_\alpha+A_hn_h
\label{eq:bcons}
\end{equation}

\noindent
and

\begin{equation}
n_e=n_p+2n_\alpha+Z_hn_h\;,
\label{eq:ccons}
\end{equation}

\noindent
where $n_n$, $n_p$, $n_\alpha$, and $n_h$ respectively denote
the number density of neutrons, protons, alpha particles, and the heavy
nucleus of mass and atomic numbers $A_h$ and $Z_h$.
In this way, the most probable configuration is found by minimizing the
total free energy of the system ${\cal F}$, subject to the above constraints

\begin{eqnarray}
\frac{\partial}{\partial x_i}\{{\cal F}
&-& V\lambda_1\left[n_B-n_n-n_p-4n_\alpha-A_hn_h\right]
\nonumber\\
&-&V\lambda_2\left[n_e-n_p-2n_\alpha-Z_hn_h\right]\}=0\;.
\label{eq:minFE}
\end{eqnarray}

\noindent
In this expression, $\lambda_i$ stands for the Lagrange-multipliers
and $x_i$ denotes $A_h$, $Z_h$, $n_n$, $n_p$, $n_\alpha$, and $n_h$.
The total Helmholtz free energy possesses the same ingredients used in
the GCA:

\begin{eqnarray}
{\cal F}&=&V[(f_n+f_n^{\rm trans})n_n+(f_p+f_p^{\rm trans})n_p\nonumber\\
&+&(f_\alpha+f_\alpha^{trans})n_\alpha+(f_h+f_h^{\rm trans})n_h]\;.
\label{eq:fetot}
\end{eqnarray}

\noindent
The contribution due to the translational motion is given by

\begin{equation}
f_i^{\rm trans}=-T\left[\log\left(\frac{g_i V_f}{\lambda_T^3}A^{3/2}\right)
-\frac{\log(k_i!)}{k_i}\right]
\label{eq:fetrans}
\end{equation}

\noindent
and, for large values of $k_i$, one may write

\begin{equation}
f_i^{\rm trans}=-T\left[\log\left(\frac{g_i\Omega}
{\lambda_T^3 n_i} A^{3/2}\right)+1\right]\;,
\label{eq:fetrans2}
\end{equation}

\noindent
since $\log(k_i!)\approx k_i\log(k_i)-k_i$.

The derivatives associated with
$n_n$ and $n_p$ allow one to easily eliminate the Lagrange multipliers:

\begin{equation}
\lambda_1=T\log\left(\frac{g_n\Omega}{\lambda_T^3n_n}\right)
\label{eq:lamb1}
\end{equation}

\begin{equation}
\lambda_2=T\log\left(\frac{n_n}{n_p}\right)
-C_c\left[-\frac{3}{2}\left(\frac{n_B}{n_0}\right)^{1/3}
          +\frac{1}{2}\left(\frac{n_B}{n_0}\right)\right]\;.
\label{eq:lamb2}
\end{equation}

\noindent
The remaining derivatives, with respect to $A_h$, $Z_h$, $n_\alpha$,
and $n_h$, lead to non-linear equations, which must be solved numerically.
Due to the logarithmic factors entering in the formulae, it is convenient to
use

\begin{equation}
n_i=\frac{g_i\Omega}{\lambda_T^3}A_i^{3/2}e^{-(f_i-\mu_i)/T}\;,
\label{eq:rhosn}
\end{equation}

\noindent
so that the equations are solved for the chemical potentials of each
species.  We require the same precision on the constraints
(\ref{eq:bcons}-\ref{eq:ccons}) as in the grand canonical
calculations.

\subsection{\label{sec:singNuc2}
A Single Nucleus Approximation with Interactions (ISNA)}

This improved single nucleus model (ISNA)
uses the mass formula from \citet{Steiner08}, adapted slightly to
include finite temperature, alpha particles, and the effects of
the neutron skin. In this model, the binding energy of a nucleus at temperature $T$,
with proton number $Z$ and mass number $A$,
in the presence of an external proton gas of density $n_{p,\mathrm{out}}$, is given by
\begin{eqnarray}
&&B(Z,A,n_n,n_p,n_{p,\mathrm{out}},T) = B_{\mathrm{bulk}}(n_n,n_p,T)
\nonumber \\
&& + \sigma {\cal B} (n_n,n_p,T)
\left(\frac{36 \pi A^2}{n^2 }\right)^{1/3} \nonumber \\
&& + {\cal  C}~\varepsilon_{\mathrm{Coulomb}}
(Z,A,n_{p,\mathrm{out}})/(An).
\label{eq:massform}
\end{eqnarray}
\noindent Here, $n_n$, $n_p$ and $n = n_n + n_p$ denote the average internal neutron, proton and baryon number densities, respectively.

The binding free energy
of bulk matter, $B_{\mathrm{bulk}}$ is given by
\begin{eqnarray}
B_{\mathrm{bulk}} &=& \frac{A-N_{\mathrm{skin}}}{n}
\left[\varepsilon(n_n,n_p,T) - n_n m_n - n_p
m_p \right. \nonumber \\
&& \left. - T s(n_n, n_p,T)\right]
+ \frac{N_{\mathrm{skin}}}{n} \left[\varepsilon(n_n,0,T) - n_n m_n
\right. \nonumber \\
&& \left. - T s(n_n, 0, T)\right]
\label{eq:bulk}
\end{eqnarray}
where $m_n$ and $m_p$ are the neutron and proton masses,
$\varepsilon(n_n,n_p,T)$ is the energy density of homogeneous matter
evaluated at the given neutron and proton density and temperature, and
$s(n_n,n_p,T)$ is the entropy density. The energy density of
homogeneous matter is described using the Akmal, Pandharipande, and
Ravenhall EOS \citep[APR]{Akmal98}. Finite temperature corrections to
the APR EOS are computed assuming that there are no finite temperature
corrections to the potential energy contributions similar to the
approach used in \citet{Prakash97}. The part proportional to
$N_{\mathrm{skin}}$ is a correction to the bulk energy density for
the neutron skin.

The average baryon density in equation~(\ref{eq:bulk}) is determined from
\begin{equation}
n = n_n+n_p = n_s + n_d I^2 \label{eq:n}
\end{equation}
where $I = 1 - 2 Z/A$. The parameter $n_s$ is analogous to the
saturation density of nuclear matter and is expected to be near 0.16
fm$^{-3}$. The parameter $n_d$ (which is typically negative) subsumes
the decrease in the saturation density with the isospin asymmetry.

The individual average neutron and proton number densities are given by
\begin{eqnarray}
n_n & = & n (1 + \delta) / 2 \nonumber \\
n_p & = & n (1 - \delta) / 2 \label{eq:np}
\end{eqnarray}
and the density asymmetry $\delta= 1 - 2 n_p / (n_n + n_p) $ is given
by $\delta = \zeta I $ where $\zeta$ is a constant parameter
determined by the fit to the experimental nuclear masses.

The parameter $N_{\mathrm{skin}}$ is chosen so that

\begin{equation}
A-N_{\mathrm{skin}} = \frac{4 \pi}{3} R_p^3 (n_n + n_p) \label{eq:as}.
\end{equation}
The neutron and proton radii, $R_n$ and $R_p$ are fixed by
\begin{eqnarray}
A-Z &=& \frac{4 \pi}{3} R_n^3 n_n \nonumber \\
Z &=& \frac{4 \pi}{3} R_p^3 n_p. \label{eq:radii}
\end{eqnarray}

In summary, for the bulk part of the nuclear mass formula, using $Z$
and $A$ one can compute the average neutron and proton densities using
the relations above, and the radii from equations~(\ref{eq:radii}), then
compute $N_{\mathrm{skin}}$ from equation~(\ref{eq:as}), and use the
equation of state for homogeneous matter to compute the bulk part in
equation~(\ref{eq:bulk}) above. Note that the symmetry energy contribution
to the nuclear mass is automatically counted above as part of
the bulk contribution.

The surface energy density as a function of the ``surface tension'',
$\sigma$ is
\begin{equation}
  \varepsilon_{\mathrm{surface}} = \frac{3 \sigma}{R}
\end{equation}
For the mass formula, we need the surface energy per baryon, which for
$T=0$ and $I=0$ is given by
\begin{equation}
B(T=0,I=0)_{\mathrm{surface}}/A = \frac{d \sigma}{n R}.
\end{equation}
Where the radius, $R$ is determined from
\begin{equation}
\frac{4 \pi R^3}{3} n = A
\end{equation}
so that
\begin{equation}
B(T=0,I=0)_{\mathrm{surface}}/A = \frac{3 \sigma}{n}
\left( \frac{4 \pi n}{3 A}\right)^{1/3}
\end{equation}

In general, the surface energy should be modified to ensure that the
surface energy vanishes in the limit $\delta \rightarrow 1$ as it
must. To address this issue, we follow \citet{Lattimer85} and approximate
${\cal B}(n_n,n_p,T)$ by:

\begin{eqnarray}
{\cal B}(n_n,n_p,T) &=& \eta(n_n,n_p,T) \frac{16 + b}
{\left[1/x^3 + b + 1/(1-x)^3\right]} \nonumber \\
&& \times B(T=0,I=0)_{\mathrm{surface}},
\end{eqnarray}
where $x=n_p/n$ and $b$ is a free parameter.

For $T>0$, one must consider the reduction in the surface tension of
the nucleus due to the interactions with the surrounding gas. For the
finite temperature correction, we follow \citep{Lattimer85} and approximate
$\eta(n_n,n_p,T)$ by:
\begin{equation}
\eta(n_n,n_p,T) = \left[\frac{\left(1-T^2/T_C^2\right)}
  {\left(1+a T^2/T_C^2 \right)}\right]^{5/4}.
\end{equation}
This expression is essentially identical to the SNA model except for
the presence of the factor $a=0.935 -5.1 (0.5-x)^2 -1.1 (0.5-x)^4$,
which is obtained from a fit to the results of \citet{Ravenhall72} that
allows extrapolations to neutron rich systems.
The critical temperature is also isospin dependent. Following \citet{Lattimer85},
we extrapolate $T_C$ to neutron rich matter using
$T_C = T_C(x=1/2) \sqrt{1-3.313 (0.5-x)^2 -7.362 (0.5-x)^4}$.  We also take
$T_c(x=1/2) = 20.085$~MeV as was done by \citet{Lattimer85}.

The Coulomb energy density~\citep{Ravenhall83} is
\begin{eqnarray}
\varepsilon_{\mathrm{Coulomb}}
&=& \frac{2}{5} \pi \left(n_p-n_{p,\mathrm{out}}\right)^2
e^2 R_p^2 \nonumber \\
&& \times \left[2 - \chi_p^{1/3}+\chi_p \right]
\end{eqnarray}
where $e^2$ is the usual Coulomb coupling~$\sim \hbar c/137$ and
$\chi_p = R_p^3/R_n^3$ is the volume fraction of matter occupied by
the proton sphere. The Coulomb contribution is multiplied in equation~(\ref{eq:massform}) by
a final parameter, ${\cal C}$, which takes into account
the fact that the proton density distribution has its own
surface. The parameter values are given in Table~\ref{tab:ldmpars} and
labeled LDM3.

In order to determine the composition and properties of matter in the
supernova, we minimize the free energy at a fixed density as a function of
the proton number and atomic number of nuclei, and the number density
of dripped neutrons, $n_{n,drip}$. The free energy of this matter
is given by
\begin{eqnarray}
&& f(Z,A,n_{n,\mathrm{out}},n_{p,\mathrm{out}},n_{\alpha,\mathrm{out}},T) =
\nonumber \\
&& n B(Z,A,n_{n,\mathrm{out}},n_{p,\mathrm{out}},T)/A
\nonumber \\
&& + (1-\chi) \left[ {f}(n_{n,\mathrm{out}},n_{p,\mathrm{out}},T)
\right. \nonumber \\
&& +\left.{f}(n_{\alpha,\mathrm{out}},T)
\right] + {f}_{\mathrm{Classical}}(n_{\mathrm{Nuclei}},T)
 + {f}_{\mathrm{el}} (n_e,T)
\end{eqnarray}
where $ n_{\mathrm{Nuclei}} = \chi n /A $, $\chi$ is the volume
fraction of matter occupied by nuclei, and
${f}_{\mathrm{Classical}}(n_{\mathrm{Nuclei}},T)$ is the classical
free energy from the ideal gas of nuclei, and ${f}_{\mathrm{el}}
(n_e)$ is the free energy of the electrons. For the purposes of
comparisons with the SNA and GCA models, we do not include the free
energy of the electrons in the ISNA calculations presented in this
paper. The free energy of the dripped nucleons,
${f}(n_{n,\mathrm{out}},n_{p,\mathrm{out}},T)$, is computed with the
APR EOS in the same way as the bulk contribution to the nuclear mass
formula referred to above.

The constraints of baryon and charge conservation are implemented with
\begin{eqnarray}
n_B &=& \chi n_n + \chi_p n_p + \left(1-\chi\right) n_{n,\mathrm{out}} +
\left(1-\chi_p\right) n_{p,\mathrm{out}}
\nonumber \\
&& + 4 \left(1 - \chi\right) n_{\alpha} \nonumber \\
Y_e n_B &=& \chi_p n_p +
\left(1-\chi_p\right) n_{p,\mathrm{out}} + 2 \left(1 - \chi \right) n_{\alpha}
\end{eqnarray}
One of the two constraints is used to fix $\chi$, and the other can be
used to constrain one of the parameters to the free energy,
e.g. $n_{p,\mathrm{out}}$.

\section{\label{sec:results}Results and Discussion\protect}

We utilize the models described above to compute the composition and
thermodynamic functions for matter for the densities
$10^{-5}<n_B/n_0<10^{-1}$, temperatures
$1~\mathrm{MeV}<T<3~\mathrm{MeV}$, and electron fractions, $0.2 < Y_e
< 0.4$ relevant for core-collapse supernovae, see
e.g. \citet{Janka07}.

\begin{figure}[htb]
\vspace*{1cm}
\begin{center}
\includegraphics[height=5.5cm,angle=0]{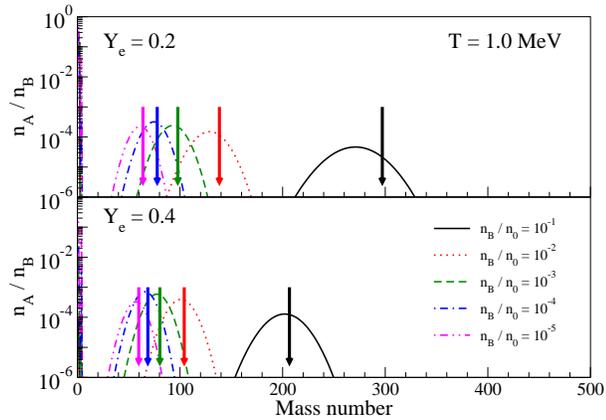}\break\break
\end{center}
\caption{\label{fig:rhoaISMM}Isobar number density for different
  values of $n_B$ at $T=1.0$~MeV, using the LDM1 mass model. The lines represent
  the results obtained with the GCA, whereas the arrows point to the
  values of $A_h$ predicted by SNA at the same values of $n_B$ used in
  the GCA. The baryon densities associated with the arrows increase
  from left to right.}
\end{figure}

The isobar number density

\begin{equation}
n_A=\sum_Zn_{A,Z}
\label{eq:nda}
\end{equation}

\noindent
is calculated through equation~(\ref{eq:rhoaz}) using the LDM1, and the
results are displayed in Figure~\ref{fig:rhoaISMM} for different baryon
densities at $T=1.0$~MeV and for two representative values of the
electron fraction used throughout this work. The lines correspond to
the distribution predicted by the GCA, whereas the arrows indicate the
value of $A_h$ in SNA. The results reveal that $A_h$ is systematically
larger than the mass number $A_{\rm max}$ at which the maximum of the
isobar number density occurs, in the heavy mass region. This may be
explained by noticing that heavy nuclei, many of them heavier than $A_{\rm
 max}$, contribute significantly to the sums in equations~(\ref{eq:rhob}) and
(\ref{eq:rhoc}) due to the smooth behavior of $n_{A,Z}$. These
contributions also include lighter nuclei such as the hydrogen isotopes (d,t), helium isotopes
($^3$He, $^6$He) and intermediate mass nuclei with $3\le Z
\le 20$. These other contributions, combined with the overall mass and charge conservation constraints, lead to $A_{\rm
  max}$ values that are lower than $A_h$. 

\begin{figure}[htb]
\vspace*{1.0cm}
\begin{center}
\includegraphics[height=5.5cm,angle=0]{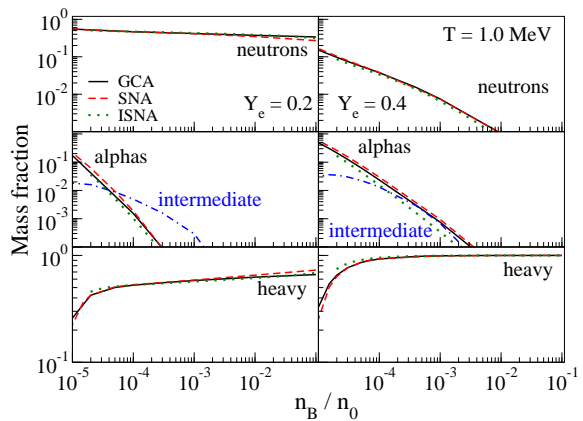}
\end{center}
\caption{\label{fig:mfracISMM} Mass fractions associated with nuclei
  with $Z=0$, $Z=2$, $3\le Z \le 20$ (intermediate),
  and $Z > 20$ (heavy), obtained employing the LDM1. The solid, dashed and dotted curves 
  denote the predictions of GCA, SNA and ISNA calculations. Upper, middle and lower panels provide the mass fractions for neutrons, alpha particles and heavy nuclei. The dashed-dotted curves in the middle left and middle right panels denote the mass fractions in the GCA calculations due to nuclei with $2\le Z\le 20$, excluding alpha particles. The left panels give results for $Y_e=0.2$, and the right panels give results for $Y_e=0.4$. All panels are for $T=1.0$~MeV.}
\vspace*{0.5cm}
\end{figure}

It may also be noticed that $n_A$ shows a broad distribution for
all baryon densities displayed in the figure. The width becomes
narrower as $n_B$ decreases, but it remains finite, for the reasons
discussed above. The position of the peak of the distribution also
shifts to lower values since dilute configurations favor
partitions with higher numbers of free nucleons
and light particles, which compete for the available charge and mass
(equations~(\ref{eq:rhob}) and (\ref{eq:rhoc})) against heavy nuclei.

This competition is illustrated in Figure~\ref{fig:mfracISMM}, which
shows the mass fraction associated with different particles. These
results clearly reveal that more mass is contained in
neutrons and alpha particles as one goes towards lower densities. The
contribution from $Z=1$ is not presented since it is relatively small for
the neutron rich systems we consider (even though it is included in 
our models). Consistent
with the conclusions of earlier work
\citep{SingleNucleus1975,BurrowsLattimer,LattimerSwesty}, the mass
fractions of alpha particles, heavy nuclei, and neutrons
predicted by the GCA and SNA calculations are very similar,
despite the non-negligible widths of the mass and charge distributions
for the CGA calculations.

Similarly, the interactions between the gas particles and between the
gas and the heavy nucleus included in the ISNA model make
comparatively small differences to the mass fractions in
Figure~\ref{fig:mfracISMM}. The main differences appear to be that the
ISNA model predicts somewhat smaller values for alpha particle and
larger values for the neutron mass fractions compared to the SNA
model. This partly reflects small differences between the mass formulae used 
in the SNA and ISNA models. In general, both models predict similar values for the the mass
of the heavy nucleus at $Y_e=0.4$, but the heavy nucleus is
somewhat lighter for the ISNA model at $Y_e=0.2$.

The qualitative dependence of $A_{\rm max}$ and $A_h$ on the electron
fraction may be understood in terms of the symmetry energy. At small
values of $Y_e$, neutrons are more abundant than protons and
consequently, the neutron chemical potential exceeds the proton
chemical potential. As a consequence, the number of free neutrons and
the neutron mass fraction increase, as can be noticed from
Figure~\ref{fig:mfracISMM}, and nuclei also tend to have more neutrons
than usual. Owing to the symmetry energy, an increase in the neutron
number increases the binding energy of protons within a nucleus, the
tendency for nuclei to have larger neutron number $N$ is accompanied
by the distribution of heavy nuclei shifting to larger $A$ and $Z$
values. The increased neutron mass fraction at $Y_e=0.2$ can only be
achieved if the heavy mass fraction correspondingly decreases. For
this reason, the mass fraction associated with heavy nuclei is smaller
for $Y_e=0.2$ compared with those obtained at $Y_e=0.4$.

\begin{figure}[htb]
\vspace*{.7cm}
\begin{center}
\includegraphics[height=6.0cm,angle=0]{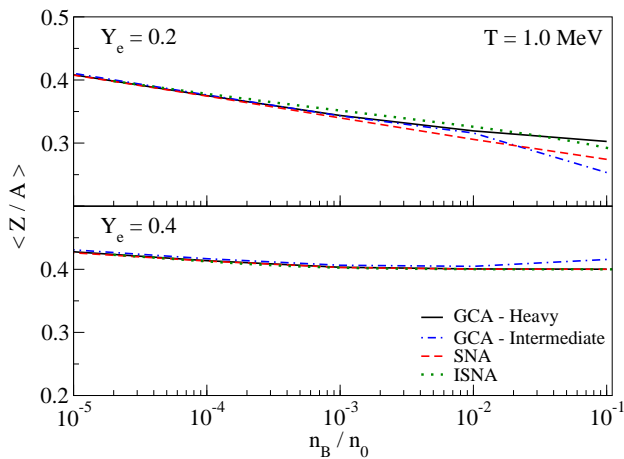}
\end{center}
\caption{\label{fig:zaISMM}Average $Z/A$ of heavy nuclei predicted by
the GCA, SNA and ISNA models, using the LDM1, are shown by the solid, dashed and dotted lines, respectively. The dashed-dotted lines show the average $Z/A$ values for lighter nuclei with $3\le Z \le 20$ that lie below the heavy nucleus charge distribution.}
\end{figure}

At higher densities, the decrease in the free volume and corresponding
increase in heavy nucleus mass means that the additional neutrons at
$Y_e=0.2$ will be largely contained in the heavy nucleus. Thus the
isotope composition of the heavy nucleus should be more sensitive to
the baryon density when the matter is appreciably asymmetric. This
aspect is illustrated in Figure~\ref{fig:zaISMM}, which shows the
average value of $Z/A$ obtained with both models. As expected, it is
fairly independent of $n_B$ for nearly symmetric matter, whereas it is
quite sensitive to it at small values of $Y_e$. The predictions of the
SNA model follow those made by the GCA fairly closely, at least at low
densities. The SNA model tends to give more neutron rich heavy nuclei,
partly because $A_h$ is systematically larger than $A_{\rm max}$ and
the isobar with the highest binding energy in nuclear mass formulae is
increasingly shifted towards the neutron rich side with increasing
nuclear mass. Some of the extra neutrons in the GCA calculations are
also contained in lighter intermediate mass nuclei with $3\le Z \le
20$ that lie below the heavy nucleus charge distribution.

The collapse of a supernova is accompanied by large changes in many
properties of the system, such as the entropy, pressure, temperature,
density, and energy \citep{Bethe1990}, some of which have an
impact on the supernova dynamics. We therefore have evaluated the entropies and energies predicted by the GCA, SNA and ISNA models. The entropy may be
obtained from these approaches through the standard thermodynamic
relation

\begin{equation}
S=\log{Z_G}+T\left[\frac{\partial}{\partial T}\log Z_G\right]
_{\mu,V}\;.
\label{eq:stherm}
\end{equation}

\noindent
Applying the relationships in the preceding section,  $S$ becomes

\begin{equation}
S=V\sum_{A,Z}n_{A,Z}\left[\frac{5}{2}+\frac{1}{T}(f_{A,Z}-\mu_{A,Z})
-\frac{\partial f_{A,Z}}{\partial T}\right],
\label{eq:sapprox}
\end{equation}

\noindent
where

\begin{equation}
\frac{\partial f_{A,Z}}{\partial T}=-5\beta_0A^{2/3}
\frac{TT_c^2}{(T_C^2+T^2)^2}
\left(\frac{T_c^2-T^2}{T_c^2+T^2}\right)^{1/4}
-\frac{2A}{\epsilon_0}T\;.
\label{eq:dfdt}
\end{equation}

\noindent
The total energy of the system is given by

\begin{equation}
E=V\sum_{A,Z}n_{A,Z}\mu_{A,Z}+TS-PV.
\label{eq:eTot}
\end{equation}

\noindent
The pressure $P$, which may be obtained from 

\begin{eqnarray}
P&=&T\left[\frac{\partial}{\partial V}\log Z_G\right]_{T,\mu}\nonumber\\
&=&
\frac{T}{\Omega}\sum_{A,Z}n_{A,Z}
\left[1+\frac{1}{2}\frac{\Omega}{T}\frac{Z^2}{A^{1/3}}C_c u\right]
\label{eq:ptherm}
\end{eqnarray}

\noindent
where

\begin{equation}
u=-\left(n_B/n_0\right)^{1/3}
+\left(n_B/n_0\right)\;.
\label{eq:upc}
\end{equation}

\noindent
The second term in equation~(\ref{eq:ptherm}) is due to the Coulomb interaction
between the electron gas and the nuclei and, therefore, is always negative
for $n_B<n_0$.

\begin{figure}[htb]
\vspace*{0.5cm}
\begin{center}
\includegraphics[height=5.5cm,angle=0]{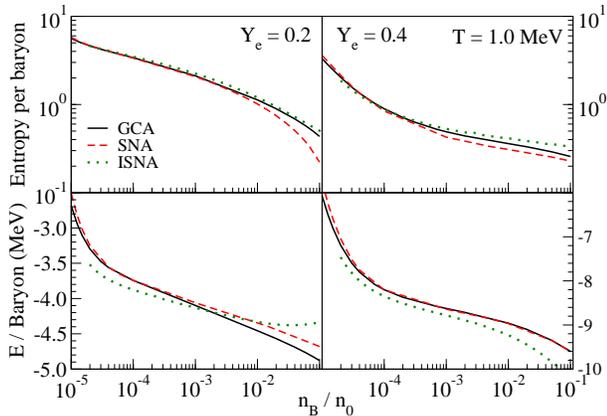}\break\break
\end{center}
\caption{\label{fig:sEtotISMM}Entropy and energy per baryon for
different values of the baryon density, obtained using the LDM1. The solid, dashed and dotted lines denote the CGA, SNA and ISNA calculations respectively. The left panels are for $Y_e=0.2$ and the right panes are for $Y_e=0.4$.}
\vspace*{0.5cm}
\end{figure}

The entropy and energy predicted by the GCA and SNA models are shown
by the solid and dashed lines in Figure~\ref{fig:sEtotISMM} for the
same baryon densities and temperature used above. We do not show the
nuclear pressure since, at these densities, it is much smaller than
that of the electron gas \citep{Bethe1990,Cooperstein1985} and,
therefore, does not play a significant role. Although a fairly
reasonable agreement is found for low densities $n_B<10^{-3}$,
differences between GCA and SNA models can be found at higher
densities where the GCA mass distributions become rather broad. The
deviations between the SNA and ISNA results for the energy and entropy
are relatively small, only a few tenths of an MeV for $Y_e=0.4$.
However, there are larger deviations at $Y_e=0.2$ where the nuclei are
extremely neutron rich. 

The differences at higher densities and low
$Y_e$ are not surprising: this is the region where one expects the
results to depend sensitively on the details of the mass formula and
information about the nuclear symmetry energy~\citep{Steiner08}.
However, the differences at low density have a different origin: near
$Y_e=0.2$ and $n_B/n_0=10^{-5}$, the free energy is very flat in the
direction of the parameter space which determines the number of nuclei
relative to the number of alpha particles. The reason for this is that
the free energy can be minimized in one of two ways: (i) the system
can choose to create more entropy (and thus decrease the free energy)
by making a lot of alpha particles from nuclei, or (ii) the system can
choose to create nuclei from alpha particles because the extra binding
created by nuclei will also decrease the free energy. Within the ISNA
model one can vary the number density of nuclei by 70 \%, and this
modification only changes the free energy by 0.1 \%, while changing
the entropy by 10 \%. Because the free energy is so flat, the entropy
in this region is quite sensitive to the nuclear mass formula.
Fortunately this low-density, low-$Y_e$ region is not actually probed
frequently in actual supernova simulations.

\begin{figure}[htb]
\vspace*{0.9cm}
\begin{center}
\includegraphics[height=5.5cm,angle=0]{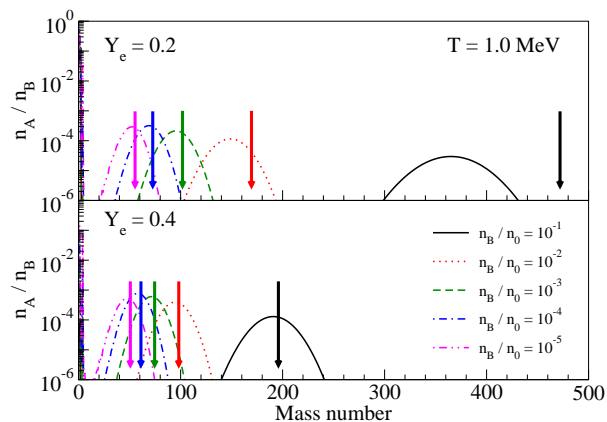}
\end{center}
\caption{\label{fig:rhoaSMF}Same quantities as in Figure~\ref{fig:rhoaISMM}.
In this figure, however, the results are obtained using the LDM2 mass model.}
\end{figure}

In order to verify the extent to which our conclusions depend on treatment of the nuclear symmetry energy in the liquid drop formula used in these statistical treatments, we have also
carried out GCA and SNA calculations using the LDM2 formula presented
in  Sect.\ \ref{sec:groundState}. The corresponding isobar number densities
are shown in Figure~\ref{fig:rhoaSMF}. Qualitatively, the calculations
in Figure~\ref{fig:rhoaSMF} resemble those in
Figure~\ref{fig:rhoaISMM} obtained using the LDM1 formula.
Quantitatively, there are differences, however. Most notably, the
discrepancies between the GCA and SNA model predictions are now much
larger for LDM2 than for LDM1, particularly for $Y_e=0.2$, at $n_B/n_0
= 0.1$. This is due to the lack of a surface symmetry energy
term in the LDM2 mass model. The resulting under prediction of the binding energies for light
masses in the LDM2 formulae shifts mass of the maximum $A_{\rm max}$ for the GCA distribution towards larger
values, where the difference between $A_{\rm max}$ and $A_h$ values is typically much larger. Nevertheless, the mass fractions predicted using the LDM2 mass model 
(not shown) are similar to those shown in Figure~\ref{fig:mfracISMM}, except
that intermediate mass nuclei ($3 \le Z\le 20$) give a somewhat larger
contribution for the LDM2 to the mass fraction at the lowest density.

\begin{figure}[htb]
\vspace*{1.0cm}
\begin{center}
\includegraphics[height=5.5cm,angle=0]{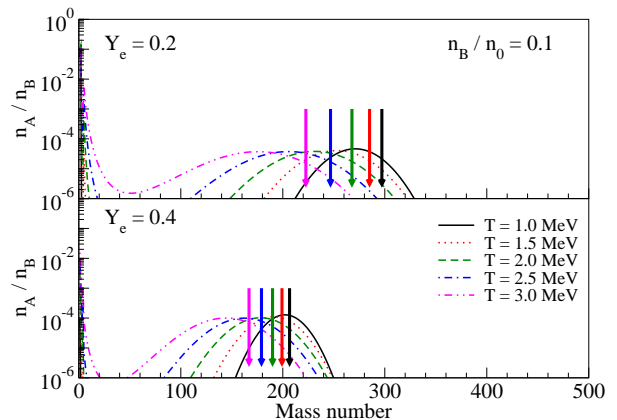}
\end{center}
\caption{\label{fig:rhoaISMM_T}Same as Figure~\ref{fig:rhoaISMM},
except that the baryon density is fixed and the temperature
is varied from 1 to 3~MeV.}
\vspace*{0.5cm}
\end{figure}

The discrepancies between the predictions of the treatments are more
pronounced at the highest baryon density. To illustrate the dependence on
temperature there, we compare the two calculations at $n_B/n_0=0.1$
for temperatures ranging from 1 to 3~MeV, values that are in the range
relevant to supernova studies. The isobar number densities, obtained
using the LDM1, are shown in Figure~\ref{fig:rhoaISMM_T}. Reflecting
the larger phase-space accessible to the system at higher
temperatures, the GCA isobar distributions become broader, the value
of $A_{\rm max}$ decreases, and a larger fraction of the mass is in
the form of nucleons or light nuclei at higher temperatures. The trend
of increasing nucleon yields with temperature is also exhibited by the
SNA model, but the discrepancies between the GCA and SNA approaches
become larger as $T$ and the widths of the mass distributions
increase. In spite of the larger width of the isobar distributions,
however, the mass fractions obtained with GCA and the SNA model (not
shown) agree to within $15 \%$ for $Y_e=0.4$ and to within about {$25
\%$} for $Y_e=0.2$. The intermediate mass nuclei have a slightly larger mass fraction, but not
enough to significantly change the mass fraction associated with heavy
nuclei.

\begin{figure}[htb]
\vspace*{1.0cm}
\begin{center}
\includegraphics[height=5.5cm,angle=0]{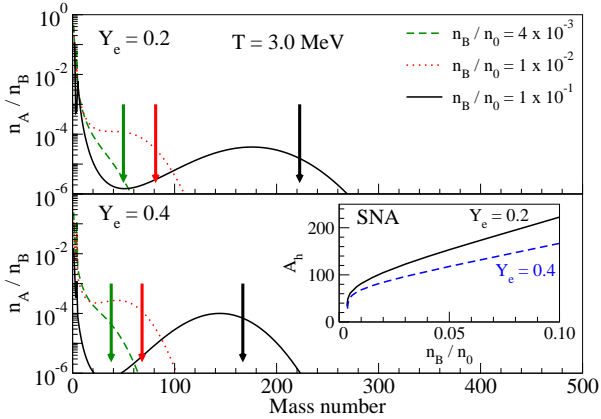}\break
\end{center}
\caption{\label{fig:rhoaISMM_T3}Same as Figure~\ref{fig:rhoaISMM_T},
  except that the $T=3$~MeV. The inset shows values for $A_{\rm h}$ calculated within
  the SNA model as a function of the density.}
\end{figure}

Figure~\ref{fig:rhoaISMM_T3} shows the complementary comparison
between the GCA and SNA calculations using the LDM1 as a function of
density at $T=3~\mathrm{MeV}$. We do not compare the two calculations 
below $n/n_0=4\times 10^{-3}$ because at this temperature the mass  $A_h$ of the most probable heavy nucleus drops to zero below this density. The inset in the figure shows the detailed
dependence of $A_h$ predicted by the SNA model. Even though, some
heavy nuclei are predicted by the GCA model at $n/n_0<4\times
10^{-3}$, the mass fraction of $Z>20$/$A>50$ at such densities is small
and of the order of $3-6\times 10^{-2}$. 
Both Figure~\ref{fig:rhoaISMM_T} and
Figure~\ref{fig:rhoaISMM_T3} show that the  discrepancies
between the GCA and SNA calculations increase with density over the range of densities
investigated in this work.

\begin{figure}[htb]
\vspace*{.7cm}
\begin{center}
\includegraphics[height=5.5cm,angle=0]{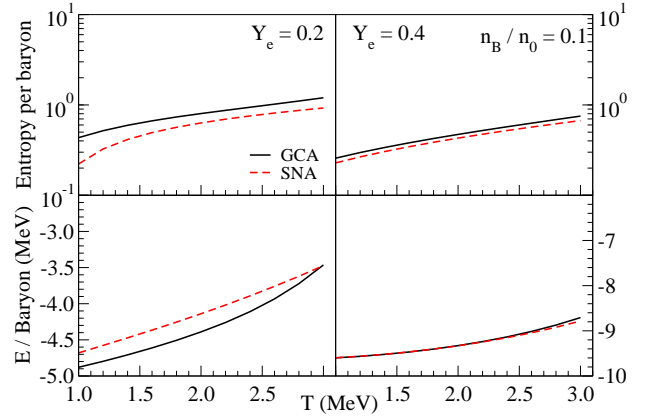}
\end{center}
\caption{\label{fig:sEtotISMM_T}
Same quantities as plotted in Figure~\ref{fig:sEtotISMM}, except that
the baryon density is fixed at $\rho/\rho_{0}=0.1$ and the temperature is varied from 1 to 3~MeV.}
\vspace*{0.5cm}
\end{figure}

In general, the agreement between the GCA and SNA approaches is
somewhat better at $Y_e=0.4$ than at $Y_e=0.2$ where the mass
distributions are wider. This conclusion also holds for the
thermodynamic variables. The dependence of the entropy and energy
per baryon on the temperature is shown at a fixed density of
$n=0.1n_0$ in Figure~\ref{fig:sEtotISMM_T}. Clear differences
between the GCA and SNA calculations are predicted for $Y_e=0.2$.
There the entropy difference ($\Delta S/A\approx0.2$) remains roughly
constant as a function of temperature. The larger mass in the SNA leads to a lower energy baryon in the SNA than in the GCA calculation. The energy
difference reaches its largest value ($\Delta E/A\approx0.15$~MeV) at
$T=1$~MeV. This difference decreases with temperature to a negligible
value at $T=3$~MeV. In contrast, the differences between GCA and SNA
calculations are comparatively small for $Y_e=0.4$, reflecting the
smaller widths of the heavy mass distributions at $Y_e=0.4$.

\begin{figure}[htb]
\vspace*{1.0cm}
\begin{center}
\includegraphics[height=5.5cm,angle=0]{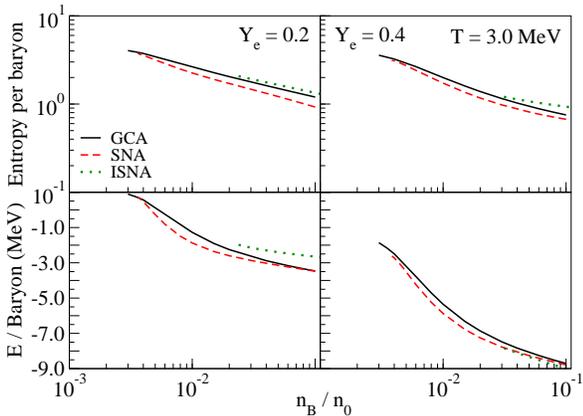}
\end{center}
\caption{\label{fig:sEtotISMM_T3} Same as Figure~\ref{fig:sEtotISMM},
except that the temperature is fixed at $T=3$~MeV and the baryon
density is varied.}
\vspace*{0.5cm}
\end{figure}

The complementary density dependencies of the entropies and energies
for GCA and SNA calculations at $T=3$~MeV are shown in Figure
\ref{fig:sEtotISMM_T3}. For both $Y_e=0.2$ and $Y_e=0.4$, the largest
differences between the energies of SNA and are actually observed at
$n/n_0 \approx0.01$. The entropy differences generally increase with
density for both $Y_e=0.2$ and $Y_e=0.4$. Both entropy and energy
differences are larger for $Y_e=0.2$ than for $Y_e=0.4$, reflecting
the larger widths of the heavy nucleus mass distributions at
$Y_e=0.2$.  Both Figs.~\ref{fig:sEtotISMM_T} and
\ref{fig:sEtotISMM_T3} demonstrate that the wide mass distributions of
the GCA calculations at densities of $n/n_0\sim0.01-0.1$ lead to
non-negligible differences between the entropies and energies
predicted by the GCA and SNA approaches. The ISNA results are similar,
again showing larger deviations from SNA calculations in the
neutron rich case at $Y_e=0.2$.  The density range over which results
are available is slightly smaller because the disappearance of the
heavy nucleus occurs as a higher density in the ISNA model. The
transition from heterogeneous to homogeneous matter, especially
in neutron-rich matter, is very model
dependent. 

\begin{figure}[htb]
\vspace*{0.9cm}
\begin{center}
\includegraphics[height=5.5cm,angle=0]{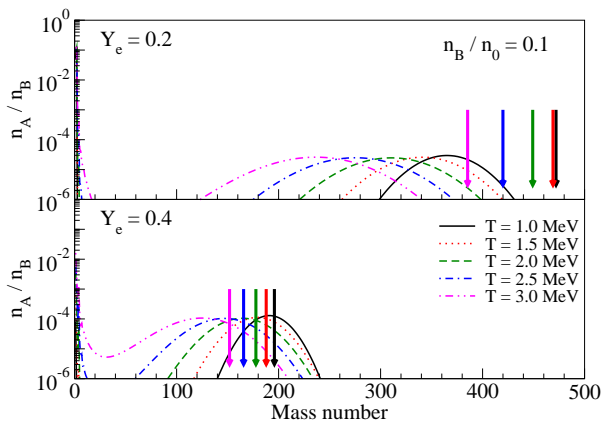}
\end{center}
\caption{\label{fig:rhoaSMF_T}
Same as Figure~\ref{fig:rhoaISMM_T}, except that
the LDM2 is used in the calculations.}
\end{figure}

Finally, to provide one more example of the importance of the mass
formula, we show, in Figure~\ref{fig:rhoaSMF_T}, the mass distributions
predicted by the GCA for the LMD2 mass formula in comparison to the
corresponding SNA predictions.  We find that there are larger
discrepancies between the GCA and SNA predicted for the LDM2
parametrization than for the LMD1 parametrization. As the mass
fractions and the thermodynamic variables obtained in this case are
very similar to those obtained for the LDM1 parametrization, we do not
show them in the interest of brevity. Comparing the predictions in two
different parameterizations of the nuclear symmetry in
Figure~\ref{fig:rhoaSMF_T}, we find that the changes in the mass
distributions due to inclusion of the surface symmetry energy
correction are less significant than the predicted differences between
the GCA and SNA calculations that use the same liquid drop mass model
(and same symmetry energy).

In principle, nuclei must have a surface symmetry energy term. The
magnitude of the surface symmetry energy coefficient, however, is not
well constrained by the measured nuclear masses
\citep{PawelMassFormula}. Even though the differences between the GCA
and SNA calculations exceed the differences between the LDM1 and LDM2
calculations, parameterizations for the density dependence of the
symmetry energy may be chosen that predict much larger effects
\citep{BotvinaSNova2005}. Ideally, a general investigation of the role
of the symmetry energy in the supernovae EOS should include both an
assessment of the sensitivity of the mass distributions to the choice
of the symmetry energy as well as an assessment of whether the chosen
form is consistent with known experimental information
\citep{isoMassFormula2008}. A systematic investigation of the
effect of the symmetry energy in relationship to the current
experimental information is in preparation.

\section{\label{sec:conclusions}Concluding Remarks\protect}

This work has focused on the differences between the Grand Canonical
Approximation and simplified Single Nucleus Approximation utilized in
most supernova calculations. Consistent with the earlier work in
\citet{BurrowsLattimer}, we find that the thermodynamic functions are
relatively well described by the SNA at densities below $0.01\rho_0$.
At higher densities, differences emerge that can be traced to the
large widths of the GCA mass distributions. We have compared
the simple SNA model to a more elaborate ISNA model that considers
effects that become relevant at higher densities, such as the
interactions between gas particles. We find differences between the
SNA and ISNA models of these average quantities that are comparable to
the differences between the GCA and SNA calculations. This indicates
the importance of taking such effects into account accurately.
Comparisons of the GCA, SNA and ISNA models suggest that the error
made by assuming a single nucleus in the energies and entropies per
baryon may be comparable to or smaller than the other uncertainties in
the problem, such as the nuclear mass formulae and the inclusion of
quantum statistics and interactions between gas particles.

Large differences can be found between the GCA predictions for the the
masses of nuclei and those of the SNA and ISNA calculations. Single
nucleus approximations, in general, tend to predict much larger heavy
nucleus masses than does the GCA. Considering first the differences
between GCA and SNA calculations, we find that the SNA model
systematically overpredicts the mass number of nuclei that are
actually present in the more accurate GCA model because it neglects
the contribution from lighter nuclei and the large width of the mass
distribution. This discrepancy grows with increasing temperature. Such
a large difference in nuclear distributions is significant. Even
though some regimes (particularly higher densities and lower
temperatures) are affected by quantum statistics, our calculations
indicate that the effects of assuming a distribution of nuclei rather
than a sole representative nucleus are much larger and sufficiently
robust. This suggests that first explorations of the consequences on
supernova simulations of these differences using classical models of
non-interacting nuclei may be helpful and should be performed.

These results has specific implications for the equation of state
tables and the weak interaction physics inputs used as inputs for
supernova simulations. Our work suggests that tables utilizing the SNA
can be updated to include nuclear distributions by applying an
appropriate GCA model. We are exploring such issues with the aim of
determining their impact on the weak interaction rates~\citep{Janka07}
in supernova simulations. The use of a GCA model will also allow
future investigations to more accurately determine the neutrino
spectra because it directly handles light nuclei, the importance of
which was emphasized in \citet{Arcones08}.

\acknowledgments

The authors would like to thank E. Brown for useful discussions
related to this work. SRS and RD were partially supported by CNPq,
FAPERJ, and the PRONEX program under contract No E-26/171.528/2006.
WGL was supported in part by the National Science Foundation under
Grant Nos.~PHY-0606007 and INT-0228058. AWS is supported by the Joint
Institute for Nuclear Astrophysics at MSU under NSF-PFC grant PHY
02-16783 and by NASA under grant NNX08AG76G. MAF acknowledges support
from the National Science Foundation under grants OISE-0735989 and
PHY-0757257.

\bibliographystyle{apj}
\bibliography{singleNucleus10_5}

\end{document}